# Formation of Embryos of the Earth and the Moon from a Common Rarefied Condensation and Their Subsequent Growth[1]

S. I. Ipatov*

*Vernadsky Institute of Geochemistry and Analytical Chemistry, Russian Academy of Sciences, Moscow, 119991 Russia *e-mail: siipatov@hotmail.com*

**Abstract** Embryos of the Moon and the Earth may have formed as a result of contraction of a common parental rarefied condensation. The required angular momentum of this condensation could largely be acquired in a collision of two rarefied condensations producing the parental condensation. With the subse-quent growth of embryos of the Moon and the Earth taken into account, the total mass of as-formed embryos needed to reach the current angular momentum of the Earth–Moon system could be below 0.01 of the Earth mass. For the low lunar iron abundance to be reproduced with the growth of originally iron-depleted embryos of the Moon and the Earth just by the accretion of planetesimals, the mass of the lunar embryo should have increased by a factor of 1.3 at the most. The maximum increase in the mass of the Earth embryo due to the accumulation of planetesimals in a gas-free medium is then threefold, and the current terrestrial iron abundance is not attained. If the embryos are assumed to have grown just by accumulating solid planetesimals (without the ejection of matter from the embryos), it is hard to reproduce the current lunar and terrestrial iron abundances at any initial abundance in the embryos. For the current lunar iron abundance to be reproduced, the amount of matter ejected from the Earth embryo and infalling onto the Moon embryo should have been an order of magnitude larger than the sum of the overall mass of planetesimals infalling directly on the Moon embryo and the initial mass of the Moon embryo, which had formed from the parental condensation, if the original embryo had the same iron abundance as the planetesimals. The greater part of matter incorporated into the Moon embryo could be ejected from the Earth in its multiple collisions with planetesimals (and smaller bodies).

*Keywords:* embryos of the Moon and the Earth, formation of the Moon, collision of rarefied condensations, angular momentum, planetesimals, lunar iron abundance, multi-impact model

**DOI:** 10.1134/S0038094618050040

## INTRODUCTION

Several models of formation of the Moon have been proposed. Its formation from a swarm of small bodies is considered in the coaccretion theory (see, e.g., Ruskol, 1960, 1963, 1971, 1975). The primary source of the near-Earth swarm of bodies in the Schmidt–Ruskol–Safronov model is the capture of particles of the preplanetary disk during their collisions ("free–free" and "free–bound"). Svetsov et al. (2012) have noted that this approach predicts the formation of satellite systems with a total mass of just ~$10^{-5}$–$10^{-4}$ of planetary mass $m_p$. In order to model the formation of massive ($0.01 m_p$–$0.1 m_p$) planetary satellites, the authors have examined the role of the material ejected in collisions between planetesimals and the Earth in the replenishment of the protolunar swarm. Svetsov et al. (2012) have concluded that the total mass of bodies ejected from the Earth in collisions between planetesimals and the Earth with a velocity of 12–20 km/s is sufficient to form a Moon-sized satellite. The hypothesis of multiple collisions (macroimpacts) between planetesimals and the Earth embryo (multi-impact model) has also been considered by Ringwood (1989), Vityazev and Pechernikova (1996), Gorkavyi (2004, 2007), Citron et al. (2014), and Rufu and Aharonson (2015, 2017). In the calculations of Citron et al. (2014), the collision velocities varied from $v_{par}$ to $1.4 v_{par}$, where $v_{par}$ is the parabolic velocity on the surface of the Earth embryo. It was demonstrated that the ratio of the mass of ejected matter to the mass of matter incorporated into the disk near the proto-Earth and the concentration of iron in the matter incorporated into the disk increase with the collision velocity. Rufu and Aharonson (2017) have demonstrated that near-vertical collisions result in lower fractions of the impactor material in the ejected matter.

It was assumed in numerous studies (e.g., Hartmann and Davis, 1975; Cameron and Ward, 1976; Canup and Asphaug, 2001; Canup, 2004, 2012;

---

[1] Reported at the Sixth International Bredikhin Conference (September 4–8, 2017, Zavolzhsk, Russia).

Canup et al., 2013; Cuk and Stewart, 2012; Cuk et al., 2016; Barr, 2016) that the Moon has formed as a result of ejection of the silicate mantle of the Earth in its col-lision with a Mars-sized body. Several modifications of the massive impact (megaimpact) model have been proposed in order to reproduce the current composition of the Earth and the Moon. Cuk and Stewart (2012) have demonstrated that a body with a mass of $(0.026–0.1)M_E$, where $M_E$ is the mass of the Earth, infalling onto the rapidly-rotating (with a period of ~2.5 h) proto-Earth may produce a lunar-forming disk consisting primarily of the terrestrial mantle material. Canup (2012) has demonstrated that the Earth and the Moon with similar compositions could be produced in a head-on collision between two bodies of similar masses (with a mass ratio no larger than 1.5). The models of Cuk and Stewart (2012) and Canup (2012) require the subsequent removal of a fraction of the angular momentum of the Earth–Moon system through the orbital resonance between the Sun and the Moon.

The semimajor axis of orbit of the formed Moon embryo in (Salmon and Canup, 2012; Cuk and Stewart, 2012) was $6r_E–7r_E$, where $r_E$ is the radius of the Earth. Owing to tidal interactions, the Earth–Moon distance may increase relatively rapidly to $30r_E$ (Touma and Wisdom, 1994; Pahlevan and Morbidelli, 2015). Pahlevan and Morbidelli (2015) have found that the Earth–Moon distance had increased to $20r_E$–$40r_E$ in $10^6$–$10^7$ years. The current Earth–Moon distance is $60.4r_E$. It was noted in (Rufu and Aharonson, 2017) that tidal interactions make the formed smaller satellite of the Earth move away from the Earth and eventually approach the more massive satellite that has been formed earlier and has originally been the more remote one.

Stewart et al. (2013) have noted that the K/Th ratios for Mercury, Venus, Earth, and Mars are similar, but this ratio for the Moon is roughly ten times lower. The low lunar K/Th ratio is attributed to the high-temperature formation in a massive impact. Norris and Wood (2017) attribute the deficit of volatiles on the Earth to the evaporation in a megaimpact and the subsequent recondensation of matter in the absence of nebular gas. Several other studies focused on the megaimpact model have been discussed in (Barr, 2016).

According to (Kaib and Cowan, 2015), the probability for the proto-Earth and the impactor to have the same oxygen isotope ratios as the current Earth and Moon is no higher than 5%. Ipatov (1993, 2000) has modeled numerically the evolution of disks of gravitating bodies merging in collisions. In the examination of the feeding zone of terrestrial planets, the initial bodies were classified into four groups according to the distance from the Sun. The simulated evolution of this disk revealed strong mixing of planetesimal bodies, and the compositions of formed planets with $m_{pl} > 0.5M_E$ were almost the same. Therefore, a considerable number of celestial bodies with similar compositions could be present in the feeding zone of the Earth and Venus (if each of these bodies formed as a result of a sufficiently large number of planetesimal collisions). The O isotope composition on the Earth varies from that on Mars, Vesta, and the majority of meteorites (Elkins–Tanton, 2013). This may be attributed to the influence of bodies beyond the Jovian orbit on the formation of Mars and asteroids. The composition of celestial bodies formed in the terrestrial region was probably more uniform and differed from the composition of Mars and asteroids. In our view, a model of the formation of the Earth and the Moon with the accumulation of a large number of planetesimals has a fair chance to reproduce the similar isotopic compositions of the Earth and the Moon.

The canonic model of a massive impact (megaimpact) has certain drawbacks of a primarily geochemical nature. It does not provide a satisfactory explanation for the compositional similarity (e.g., the closeness of concentrations of isotopes of oxygen, iron, hydrogen, silicon, magnesium, titanium, potassium, tungsten, and chromium) of the Earth and the Moon, since the greater part of lunar matter in this model originates from the impactor instead of from the proto-Earth (Galimov et al., 2005; Galimov, 2011; Galimov and Krivtsov, 2012; Elkins–Tanton, 2013; Clery, 2013; Barr, 2016). It is assumed in the megaimpact hypothesis that a magma ocean forms on the planetary surface after the collision. Jones (1998) has noted the lack of evidence in favor of the existence of such an ocean on the Earth at any point in history. According to Galimov (2011), the megaimpact theory fails to account for the lack of an isotope shift between lunar and terrestrial matter because the ejected material should be 80–90% vapor, and the K, Mg, and Si isotope compositions may change considerably during evaporation.

A model of formation of embryos of the Moon and the Earth as a result of contraction of a rarefied dust condensation in a protoplanetary gas–dust cloud has been proposed in the work of (Galimov et al., 2005; Galimov, 1995, 2008, 2011, 2013; Galimov and Krivtsov, 2012; Vasil'ev et al., 2011). The evaporation of FeO from dust particles is taken into account in this model, which agrees better with the geochemical data on the composition of lunar matter. The authors of the above studies have noted that the hypothesis of parallel formation of the Moon and the Earth in the collapse and fragmentation of a large dust condensation agrees with geochemical evidence.

It follows from the analysis of the $^{182}$Hf–$^{184}$W system performed by Galimov (2013) that the Moon could not have formed earlier than 50 million years after the origin of the Solar System. Having studied the Rb–Sr system, Galimov concluded that the Moon should have evolved in a medium with a higher Rb/Sr ratio prior to its emergence as a condensed body. The large atomic weight of rubidium makes its escape from the lunar surface impossible; this may occur only on the heated surface of small bodies or particles. Therefore, in Galimov's view, the initial lunar matter remained in a dispersed state (e.g., in the form of a gas–dust condensation) for the first 50 million years. It was assumed in the above-mentioned papers by Galimov and his coauthors that the stability of a gas–dust condensation could be maintained for a considerably long time by intense gas emission from the surface of particles and, possibly, by ionization and radiation repulsion due to the decay of short-lived isotopes. Note that the protosolar gas–dust cloud had existed before its contraction and the formation of the Sun and the protoplanetary disk, and rubidium could also escape from particles, e.g., during the formation of the Sun.

Galimov et al. (2005), Galimov (2011), and Galimov and Krivtsov (2012) have modeled numerically the formation of embryos of the Earth–Moon system from a rarefied condensation and have studied the growth of solid embryos of the Moon and the Earth by particle accumulation. Galimov and Krivtsov (2012) have reported the results of calculations of contraction of a condensation with a mass equal to the combined mass of the Moon and the Earth and a radius of $5.5r_E$. Galimov et al. assumed that ~40% of volatile matter (FeO included) of dust particles, which formed the embryos, evaporated, and the initially high-temperature embryos of the Moon and the Earth were similarly depleted in iron. This 40% reduction in the particle mass transpired within $(3–7)\times10^4$ years in their model. According to the estimates of Galimov, the evaporation of 40% of the mass of matter of the initial chondritic composition results in a reduction in the concentration of iron to lunar levels. The evaporation of material from the surface of particles made the interval of condensation contraction longer. Other particles in the condensation, which were located at a greater distance from its center and were not incorporated into the original embryos, were cooler and retained iron. The vapor flow from the surface of particles produced a repulsive force that, acting together with gas, prevented the contraction of the condensation. The embryos grew and accumulated iron-enriched particles that were located in the outer part of the condensation at the moment of embryo formation. The embryo of the Earth grew faster than the Moon embryo. This is the reason why the Moon has retained a relatively low iron abundance, while the Earth has accumulated the greater part of the remaining dust condensation and acquired a considerable amount of iron. Vasil'ev et al. (2011) and Galimov and Krivtsov (2012) have modeled the collisions between particles and embryos. The initial positions of dust particles were distributed uniformly over a cylindrical surface, and the initial velocities were zero. The authors have found that a 26.2-fold increase in the mass of the Earth embryo corresponds to a 1.31-fold increase in the mass of the Moon embryo.

Galimov and Krivtsov (2012) have demonstrated that their model of formation of the Earth and the Moon agrees with geochemical data. For example, the formation of the Earth and the Moon from a common condensation explains why the O isotope composition ($^{16}O/^{17}O/^{18}O$) and the $^{53}Cr/^{52}Cr$, $^{46}Ti/^{47}Ti$, and $^{182}W/^{184}W$ ratios are the same on the Moon and the Earth. In view of Galimov and Krivtsov (2012), the similarity of isotopic characteristics of the Earth and the Moon presents unsurmountable problems for the megaimpact hypothesis. These authors have also demonstrated that their model provides a much better (com-pared to the megaimpact hypothesis) explanation for the following data: (1) the abundance of siderophile elements (W, P, Co, Ni, Re, Os, Ir, Pt, etc.) on the Earth and in the lunar mantle is lower than the expected values based on the known distribution coefficients; (2) Hf/W isotope data for the modern Earth and Moon; (3) isotopic geochemistry of Xe, Pb, and Sr. Galimov et al. (2005) have noted that the observed distribution of siderophile elements on the Moon could also be obtained from the initial material, and its core has formed in the conditions of partial melting.

According to the model proposed in (Galimov and Krivtsov, 2012), ~50–70 million years after the beginning of formation of the Solar System, a rarefied condensation with a mass equal to that of the Earth–Moon system contracted in $10^4–10^5$ years and thus formed embryos of the Moon and the Earth. Such long existance times of condensations in the early Solar System have not been obtained by scientists specializing in the formation and evolution of condensations. Marov et al. (2008) believe that the evolution of the circumsolar protoplanetary disk to the point of formation of a dust-enriched subdisk took 1–2 million years, and the subdisk then contracted and formed dust condensations within ~0.1 million years. In the model of Makalkin and Ziglina (2004), trans-Neptunian objects with diameters up to 1000 km form within an interval on the order of a million years after the onset of formation of the Solar System. In the majority of recent studies focused on the formation of planetesimals (Cuzzi et al., 2008, 2010; Cuzzi and Hogan, 2012; Johansen et al., 2007, 2009a, 2009b, 2011, 2012; Lyra et al., 2008; 2009; Youdin, 2011; Youdin and Kenyon, 2013), the actual time of formation (after the onset of formation from a condensed gas–dust disk) and contraction of rarefied condensations does not exceed 1000 revolutions about the Sun; in certain models, it is as short as several tens of revolutions about the Sun. The contraction of condensations and the formation of satellite systems in the trans-Neptunian belt occur within ~100 years in the model of Nesvorny et al. (2010). In order to obtain longer contraction times, one needs to take the factors inhibiting the process of contraction of rarefied condensations into account. The times of contraction of condensations to the density of solid bodies in the studies of Myasnikov and Titarenko (1989a, 1989b) are as long as several million years (depending on the optical properties of dust and gas and the type and concentration of short-lived radioactive isotopes in condensations). Beletskii and Grushevskii (1991) have found that the angular momentum of contracting rarefied protoplanets could decrease considerably due to tidal interactions with the Sun.

Several authors consider the formation of condensations with masses exceeding that of Mars possible. For example, the formation of rarefied condensations with a mass of ~$0.1M_E$–$0.6M_E$ was examined in (Lyra et al., 2008). These condensations form due to the Rossby wave instability rather than by accumulating multiple smaller condensations. Ipatov (2017a) has reviewed the studies focused on the formation of rarefied condensations.

In the present study, the possible scenarios for the formation of embryos of the Moon and the Earth from a rarefied condensation and the subsequent growth of these embryos are discussed. The concept of contraction of rarefied condensations, which was considered earlier by Ipatov (Ipatov, 2010, 2014; Ipatov, 2017b, 2017c) and Nesvorny et al. (2010) in the context of the formation of trans-Neptunian satellite systems, is taken as a basis. Ipatov (2010) and Nesvorny et al. (2010) have assumed that trans-Neptunian satellite systems have formed as a result of contraction of rarefied condensations. Ipatov (2010) has demonstrated that the angular momenta of the observed trans-Neptunian satellite systems are equal to the angular momenta of colliding rarefied condensations of the same masses smaller than their Hill spheres. The angular momentum of two colliding condensations may be negative, which is also true of the angular momentum of trans-Neptunian satellite systems. Nesvorny et al. (2010) have calculated the contraction of rarefied condensations in the trans-Neptunian region and found the initial conditions in which this contraction resulted in the formation of binary (or triple) systems. They have found that the gas resistance forces do not exert any significant effect on the formation of a binary system via contraction. Ipatov (2017b) has demonstrated that the angular momentum needed to form trans-Neptunian satellite systems in the process of contraction of parental condensations could be acquired in condensation collisions. This model of formation of trans-Neptunian satellite systems may provide an explanation for the observed orbits of components of these systems (Ipatov, 2017c). Ipatov (2015) has demonstrated that the angular momentum needed to form embryos of the Moon and the Earth in the process of contraction of a parental rarefied condensation could be acquired in a collision between two condensations. Ipatov (2017a) believed that the formation of these embryos was similar to the formation of trans-Neptunian satellite systems.

## 1. FORMATION OF EMBRYOS OF THE EARTH AND THE MOON AT THE STAGE OF RAREFIED CONDENSATIONS

### 1.1. Angular Momentum of a Condensation Formed in a Collision of Two Condensations

Ipatov (2010, 2014, 2017b, 2017c) has considered a model with satellite systems of small bodies forming as a result of collisions of condensations that produce a condensation with sufficient angular momentum. Ipatov (2010) has found that the angular momentum of two colliding condensations (with radii $r_1$ and $r_2$ and masses $m_1$ and $m_2$), which had circular heliocentric orbits with semimajor axes close to $a$ prior to the collision, is

$$K_s = k_\Theta (G \cdot M_S)^{1/2} (r_1+r_2)^2 \cdot m_1 \cdot m_2 \cdot (m_1+m_2)^{-1} a^{-3/2}, \quad (1)$$

where $G$ is the gravitational constant, $M_S$ is the mass of the Sun, and the difference between the semimajor axes of orbits of condensations is $\Theta(r_1 + r_2)$. At $(r_1 + r_2)/a \ll \Theta$, $k_\Theta \approx 1-1.5\Theta^2$. The $k_\Theta$ values vary from –0.5 to 1. The average value, $|k_\Theta|$, is 0.6. The values of $K_s$ and $k_\Theta$ are positive at $0 < \Theta < 0.8165$ and negative at $0.8165 \ll 1$. If two identical condensations with their radii equal to $k_H r_H$, where $r_H$ is the Hill radius of a condensation with mass $m_1 = m_2$, collide, it follows from (1) that

$$K_s \approx 0.96 k_\Theta \cdot k_H^2 \cdot a^{1/2} \cdot m_1^{5/3} \cdot G^{1/2} \cdot M_S^{-1/6} \quad (2)$$

Let us denote the angular momentum in a typical collision of two identical condensations, which are the size of their Hill spheres, in circular heliocentric orbits as $K_{s2}$. Using formula (2), one may determine that the ratio of angular momentum $K_{\Sigma EM}$ of the Earth–Moon system ($K_{\Sigma EM} \approx 3.45 \times 10^{34}$ kg m$^2$ s$^{-1}$) to angular momentum $K_{s2}$ is roughly equal to 0.0335 at $k_\Theta = 0.6$ and to 0.02 at $k_\Theta = 1$ if condensation masses $m_1 = 0.5 \times 1.0123 M_E$. Thus, the angular momentum in such a collision in the considered model may be 50 times higher than the current angular momentum of the Earth–Moon system. If the eccentricities of heliocentric orbits are nonzero, the angular momentum of colliding condensations may exceed the value for circular orbits.

In the model considered below, the mass and the angular momentum of a condensation produced in a collision are the same as those of the colliding condensations. In reality, a fraction of the mass and the angular momentum gets lost in the collision (especially in grazing collisions) and in the process of condensation contraction. Therefore, the mass and the angular momentum of colliding condensations could be larger than those of the parental condensation and the satellite system formed as a result of contraction of the parental condensation. It follows from Fig. 2 in (Nesvorny et al., 2010) that the mass of the formed solid binary object was approximately 5 times lower than the mass of the parental condensation. The introduction of the effect of gas within the condensation into calculations is likely to reduce the mass and momentum loss in the formation of embryos.

Since $K_{s2}$ is proportional to $m_1^{5/3}$, $K_{s2} = K_{\Sigma EM}$ for $k_\Theta = 0.6$ at $2m_1 \approx 0.0335^{3/5} \times 1.0123 M_E \approx 0.13 M_E$. In the case of circular heliocentric orbits, the maximum (at $k_\Theta = 1$) $K_{s2}$ value of $1.7 \times 10^{36}$ kg m$^2$ s$^{-1}$ is $0.6^{-1}$ times higher than the above typical one (at $k_\Theta = 0.6$). Then, $K_{\Sigma EM}/K_{s2} \approx 0.0335 \times 0.6 \approx 0.02$ and $2m_1 \approx 0.02^{3/5} M_E \approx 0.096 M_E$. Thus, the angular momentum of the Earth–Moon system could be acquired in a collision of two condensations in circular heliocentric orbits with their total mass being no lower than the mass of Mars.

Surville et al. (2016) have concluded that large-scale dust rings, which are then subjected to streaming instability, form after vortex dissipation. The ring mass in their models could be as large as $0.6 M_E$, and the ring width was on the order of $(2–3) \times 10^{-3} a$, where $a$ is the distance between the ring and the Sun. Such a ring makes the formation and collision of two condensations in relatively close orbits possible. It is mentioned below in Subsection 2.1 that the initial mass of a rarefied condensation producing embryos of the Earth and the Moon may be relatively small ($0.01 M_E$ or even smaller) if one takes into account the increase in the angular momentum of embryos associated with the increase in their mass.

Since $K_s = J_s \omega_c$, it follows from (1) that the angular velocity of a condensation produced in a collision of two condensations is

$$\omega_c = 2.5 k_\Theta \chi^{-1} (r_1 + r_2)^2 r^{-2} m_1 m_2 (m_1 + m_2)^{-2} \Omega, \quad (3)$$

where $\Omega = (GM_S/a^3)^{1/2}$ is the angular velocity of motion of the condensation around the Sun. The moment of inertia of the condensation with radius $r$ and mass $m$ is $J_s = 0.4 \chi m r^2$, where $\chi$ characterizes the matter distribution within the condensation ($\chi = 1$ for a homogeneous spherical condensation considered by Nesvorny et al. (2010)). At $r_1 = r_2$, $r^3 = 2 r_1^3$, $m_1 = m_2 = m/2$, and $\chi = 1$, $\omega_c = 1.25 \times 2^{1/3} k_\Theta \Omega \approx 1.575 k_\Theta \Omega$.

According to Safronov (1969), the initial angular velocity of a rarefied condensation (relative to its center of mass) is $0.2\Omega$ for a spherical condensation and $0.25\Omega$ for a plane circle. The initial angular velocity is always positive and may be almost an order of magnitude lower than the angular velocity acquired in a collision of condensations. The initial angular velocity of a condensation is insufficient to form a satellite system (see below).

The contribution of the initial rotation to angular momentum $K_s$ of the parental condensation may be more significant if the condensation contracted prior to a collision. Let us consider a collision of two identical spherical condensations with masses $m_1$ and radii equal to $k_{col} r_H$ ($r_H$ is the Hill radius of the condensation). It is assumed that each of them was initially formed with a radius of $k_{in} r_H$ and an angular velocity of $0.2\Omega$. The angular momentum of a spherical condensation produced after a collision is then

$$K_s \approx (0.96 \cdot k_\Theta \cdot k_{col}^2 + 0.077 \cdot \chi \cdot k_{in}^2) a^{1/2} m_1^{5/3} G^{1/2} M_S^{-1/6}.$$

The contribution of the initial rotation at $\chi = 1$ is larger than that of the collision if $k_{in}/k_{col} > 2.7$ and $k_\Theta = 0.6$ (or $k_{in}/k_{col} > 3.5$ and $k_\Theta = 1$) in this formula. If we consider condensations that are denser at the center ($\chi < 1$), the contribution of the collision to $K_s$ may be larger than that of the initial rotation at $k_{in}/k_{col} < 3\chi^{-1/2}$.

It follows from formula (3) that $\omega_c$ is proportional to $2^{2/3} k_r^3 (1 + k_r)^2 (1 + k_r^3)^{-8/3}$ at $r_2 = k_r r_1$. Specifically, the values of $\omega_c$ at $k_r = 0.5$ and $k_r = 1/3$ are lower than the value at $k_r = 1$ by a factor of approximately 3 and 10, respectively.

Let us consider the merger of two colliding condensations of equal densities with masses $k_m m$ and $(1 - k_m) m$ where $0 < k_m < 1$ and an initial angular velocity of $k_\Omega \Omega$ (the typical $k_\Omega$ value is 0.2). The component of the angular momentum of the formed condensation with radius $r$ associated with the initial rotation of colliding condensations is then $K_{sin} = k_\Omega \Omega (0.4 \chi m r_{in}^2)[(1 - k_m)^{5/3} + k_m^{5/3}]$, where $r_{in}$ is the radius of the condensation with mass $m$ and density equal to that of the initial condensations. The collision-induced component of the angular momentum of the condensation formed with mass $m$ and radius $r_{col}$ is

$K_{sc} = k_\Theta \cdot \Omega \cdot m \cdot r_{col}^2 \cdot k_m(1 - k_m) \cdot [(1 - k_m)^{1/3} + k_m^{1/3}]^2$. At $k_\Omega = 0.2$, $k_\Theta = \chi = 1$, $r_{col} = r_{in}$, and $k_m = 1/28$, $K_{sc} \approx 0.8 K_{sin}$; therefore, if the condensations did not contract prior to the collision and the ratio of their radii is 3 (the mass ratio is 27), the contribution of the initial rotation to the final angular momentum is slightly larger than the collisional contribution. Thus, in order for the contribution of the collision of condensations to the angular momentum of the parental condensation to be more significant than that of the initial rotation, the radii of colliding condensations should decrease by a factor of no more than three prior to the collision and should differ by a factor of no more than three.

### 1.2. Angular Momentum of a Condensation Formed by Accumulation of Smaller Objects

Condensations and embryos formed from a condensation may grow by accumulating smaller objects. In the models of Drazkowska et al. (2016), planetesimals formed from condensations typically incorporated ~20% of the total amount of solid matter, while the remaining matter of the protoplanetary disk went into smaller objects. A certain fraction of the mass and the angular momentum of the parental rarefied condensation, which had contracted and formed embryos of the Earth–Moon system, could be acquired in the process of accumulation of smaller objects by the parental condensation. Ipatov (1981b, 2000, 2017b) has studied the angular momentum of a condensation for several models of its growth by accumulation of smaller objects.

If radius $r$ of a growing condensation is equal to $k_H r_H$ ($k_H$ is a constant and $r_H$ is the Hill radius of the growing condensation) and the modulus of its tangential velocity component is $|v_\tau|=0.6 v_c \cdot r \cdot a^{-1}$, angular momentum $K_s$ of a condensation with mass $m_f$, which has grown by accumulating small objects, is written as (Ipatov, 2017b)

$$K_s \approx 0.173 k_H^2 G^{1/2} a^{1/2} m_f^{5/3} M_S^{-1/6} \Delta K, \qquad (4)$$

where $v_c$ is the velocity of motion of this condensation in a circular heliocentric orbit with radius $a$, $M_S$ is the mass of the Sun, and $K = K^+ - K^-$ is the difference between positive $K^+$ and negative $K^-$ changes in the angular momentum of the condensation after the infall of small celestial objects ($K^+ + K^- = 1$). The condensation contraction in the process of accumulation of smaller objects was neglected in the derivation of (4). The values of $K$ for various eccentricities and semi-major axes of heliocentric orbits and masses of objects approaching the condensation to within the radius of the considered sphere were given in (Ipatov, 1981a, 1981b) and, in brief, in (Ipatov, 2000). It was found that $K \approx 0.9$ at almost circular heliocentric orbits of objects and a condensation radius close to the Hill radius. If the condensation growth from $m_o$ to $m_f$ is considered, $m_f^{5/3}$ in (4) should be replaced by $m_f^{5/3} - m_o^{5/3}$. Formula (4) is valid both for condensations and for solid bodies. Infalling dust particles and bodies could be originally located at different distances from the Sun away from the condensation (the longer the condensation life-time is, the farther away they could be positioned). Dust particles could migrate toward the condensation under the influence of gravity, radiation pressure, solar wind, and the Poynting–Robertson effect. The motion of bodies is influenced by gravity and the Yarkovsky effect.

### 1.3. Angular Momentum of a Condensation Needed to Form Embryos of the Earth and the Moon

The initial angular velocities of condensations were taken equal to $\omega_o = k_\omega \Omega_o$, where $\Omega_o = (Gm/r^3)^{1/2}$ is the circular velocity on the condensation surface, in the calculations of contraction of condensations (with mass $m$ and radius $r$) in the trans-Neptunian region performed by Nesvorny et al. (2010). The values of $k_\omega$ = 0.5, 0.75, 1, and 1.25 and condensation radii equal to $0.4 r_H$, $0.6 r_H$, and $0.8 r_H$, where $r_H$ is the Hill radius of a condensation with mass $m$, were considered. Note that $\Omega_o/\Omega = 3^{1/2}(r_H/r)^{3/2} \approx 1.73(r_H/r)^{3/2}$. If $r \ll r_H$, then $\Omega \ll \Omega_o$. In the case of Hill spheres, assuming that angular velocity $\omega_c \approx 1.575 k_\Theta \Omega$ of a condensation formed in a collision of two identical condensations is equal to $\omega_o$, we obtain $k_\omega \approx 0.909 k_\Theta/\chi$. This implies that one may obtain the values of $\omega_c = \omega_o$ corresponding to $k_\omega$ up to 0.909 in condensation collisions with $k_\Theta = \chi = 1$.

In the case of collision of two condensations, the size of their Hill spheres and subsequent contraction of the formed condensation to radius $r_c$, the angular velocity of the contracted condensation is $\omega_{rc} = \omega_H (r_H/r_c)^2$, where $\omega_H \approx 1.575 k_\Theta \Omega$. Assuming that $\omega_o = k_\omega (Gm/r_c^3)^{1/2}$, we find that $\omega_{rc}/\omega_o$ for such a condensation with radius $r_c$ is proportional to $r_c^{-1/2}$. At $r_c/r_H = 0.6$, for angular momentum $K_s$ of colliding condensations the size of their Hill spheres may correspond to $k_\omega$ up to $0.909/0.6^{1/2} \approx 1.17$. In (Nesvorny et al., 2010), binary or triple systems were obtained only at $k_\omega = 0.5$ or 0.75. Thus, it follows that the initial angular velocities of condensations corresponding to the formation of binary systems could be acquired in condensation collisions.

Let us compare the angular velocity acquired by a condensation while accumulating smaller objects with the angular velocity $\omega_o$ needed to form a satellite system during condensation contraction. Comparing $K_s = J_s \omega_o$ ($\omega_o = k_\omega \Omega_o$ and $J_s = 0.4 \chi m r^2$) to the value of $K_s$ calculated using formula (4), we obtain $K \approx 0.8 \chi k_\omega$ (for any $r$ and $m$). It follows that $K$ at $\chi = 1$ is roughly equal to 0.4, 0.5, and 0.6 at $k_\omega = 0.5$, 0.6, and 0.75, respectively. The variation of the condensation density and $\chi$ in the process of accumulation is neglected in these estimates. Since the density may be higher at smaller distances from the condensation center, the typical $\chi$ value is lower than unity. The $K$ values are normally lower for colliding objects with higher densities and higher eccentricities of heliocentric orbits (Ipatov, 1981a, 1981b, 2000). The above estimates do not contradict the notion that a condensation growing by accumulating smaller objects could acquire, in certain cases, an angular velocity needed to form a binary system.

Since $\Omega_o/\Omega \approx 1.73(r_H/r)^{3/2}$, $\Omega \approx 0.58 \Omega_o$ at $r = r_H$, and the initial angular velocity of rotation of a rarefied spherical condensation about its center of mass is (Safronov, 1969) $0.2\Omega \approx 0.12 \Omega_o$. If $r \ll r_H$, then $\Omega \ll \Omega_o$. It follows from the above estimates that the angular velocity and the angular momentum of a condensation acquired in the process of its formation from a protoplanetary disk were insufficient to form a satellite system.

Galimov and Krivtsov (2012) and Le-Zakharov and Krivtsov (2013) have calculated the gravitational collapse of a condensation with a mass equal to the current mass of the Earth–Moon system ($m \approx 1.01 M_E$) and radius $r \approx 5.5 r_E \approx 0.023 r_H$, where $r_E$ is the radius of the Earth. In their two-dimensional calculations, a satellite system formed at an initial angular velocity of condensation rotation $\omega_o > 0.64\Omega_o$, but an average number of two formed clusters was attained at $\omega_o \approx 1.1\Omega_o$. Galimov and Krivtsov (2012) have considered $\omega_s = (3\pi/4)^{1/2}\Omega_o \approx 1.535\Omega_o$ instead of $\Omega_o$ used in (Nesvorny et al., 2010). In the three-dimensional calculations in (Galimov and Krivtsov, 2012), two embryos formed if the angular velocity of condensation rotation fell within the interval from $\Omega_o$ to $1.46\Omega_o$. At $\omega_o < \Omega_o$, only a central body without satellites formed in most cases, and a considerable fraction of the momentum could be carried away by particles leaving the contracting condensation. In (Nesvorny et al., 2010), satellites formed at lower initial angular velocities (falling within the $0.5\Omega_o$–$0.75\Omega_o$ range). The discrepancies between the results of these two studies may be attributed to the differences in chaotic velocities of particles/bodies forming condensations, in modeling techniques, and in masses and sizes of the condensations considered. Several satellites could be formed at higher $\omega_o$ values. Galimov and Krivtsov (2012) have considered the evaporation from millimeter particles, which had formed a rarefied condensation, in order to simulate the formation of the Earth–Moon system in the process of contraction of a condensation with the same (as for this system) angular momentum ($K_s = K_{\Sigma EM}$ at $r \approx 5.5r_E$ and $\omega_o \approx 0.12\Omega_o$) and obtain embryos with low iron abundances (iron was removed partially from the particles during evaporation). In the model with evaporation, two embryos formed at $\omega_o \approx 0.12\Omega_o$. In the model without evaporation, the angular momentum for a condensation with $m = 1.0123M_E$ and $r = 0.023r_H$ at $\omega_o = \Omega_o$ is roughly eight times higher than that at $\omega_o \approx 0.12\Omega_o$ (i.e., $K_s \approx 8K_{\Sigma EM}$).

Any angular momentum values used in (Galimov et al., 2005; Galimov and Krivtsov, 2012) could be acquired in collisions of condensations with a total mass lower than the mass of the Earth. In order to acquire the needed angular momentum, the condensation produced in a collision should have a radius larger than $r = 0.023r_H$ (the value used in the calculations of the above authors), although it may be smaller than the Hill radius. The parental condensation formed in a collision may contract to $r = 0.023r_H$.

As noted in Subsection 1.1, angular momentum $K_{\Sigma EM}$ of the current Earth–Moon system could be acquired in a collision of two rarefied condensations (with their radii equal to $r_H$) in circular heliocentric orbits with total mass $m_{tot}$ no lower than $0.1M_E$. At $m_{tot} \approx 1.01M_E$ and condensation radii equal to their Hill radii, the angular momentum may be as large as $50K_{\Sigma EM}$. Therefore, even if a considerable fraction of the angular momentum is lost in the process of condensation contraction, the angular momenta considered in (Galimov et al., 2005; Galimov and Krivtsov, 2012) still remain feasible.

Angular momentum $K_{si}$ in a collision of two identical condensations with total mass $m_f = 1.0123M_E$ is equal to $K_{\Sigma EM}$ at $k_H \approx 0.17$ and $k_\Theta = 0.6$ or at $k_H \approx 0.13$ and $k_\Theta = 1$. These relations demonstrate that if the major part of the angular momentum of the parental condensation with a mass equal to that of the current Earth–Moon system was acquired in a collision of two identical condensations, their radii were larger than $0.1r_H$. This value is higher than the radius of the parental condensation ($0.023r_H$) examined in (Galimov and Krivtsov, 2012). Therefore, in order to acquire the needed angular momentum, the condensation considered in this study should be the result of the contraction of a larger condensation. At $k_H = 0.02$, we have $K_{si} = K_{\Sigma EM}$ only at $m_f \approx 13M_E$.

The above estimates suggest that any initial angular velocities and momenta considered in (Galimov and Krivtsov, 2012; Nesvorny et al., 2010) could be attained after the contraction of a condensation produced in a collision of condensations fitting within their Hill spheres.

The radii of initial condensations considered in the modeling of condensation formation are usually comparable to the Hill radii. The condensation formed after contraction of a larger condensation to a radius of $0.023r_H$ could contain objects larger than the millimeter dust particles examined in (Galimov and Krivtsov, 2012). It was demonstrated in several studies (see, e.g., Johansen et al., 2007) that condensations in the feeding zone of terrestrial planets contained decimeter-sized objects. These objects could have a fractal structure (Kolesnichenko and Marov, 2013; Marov, 2017), and the mechanism of FeO evaporation from their surface corresponded to the one considered in (Galimov and Krivtsov, 2012). Technically, the condensation with a radius of $0.023r_H$ was regarded in the studies of Galimov et al. as the central region of a condensation with a radius of $r_H$. However, the question still remains how such a massive ($1.01M_E$) small central region of a condensation could form from millimeter particles.

In the general case, the initial distance between embryos formed in the process of condensation contraction could be rather large and even close to the Hill radius (the Hill radii for $1.01M_E$, $0.1M_E$, and $0.02M_E$ are $235r_E$, $109r_E$, and $64r_E$, respectively). However, the distance between the initial embryos in the megaimpact and multi-impact models and the model of Galimov et al. was small.

In the model of condensation growth by accumulation of small objects, the $K_s$ value calculated using Eq. (4) at $K = 0.9$, $m_f = M_E + M_M$ (the sum of current masses of the Earth and the moon), $k_H = 1$, and $a = 1$ AU is more than 24.5 times higher than the current angular momentum $K_{\Sigma EM}$ of the Earth–Moon system (including the moment of axial rotation of the Earth).

Since the $K_s$ value in Eq. (4) is proportional to $m_f^{5/3}$, $K_s = K_{\Sigma EM}$ at $m_f \approx 0.15(M_E + M_M)$ and $K = 0.9$ or at $m_f \approx 0.2(M_E + M_M)$ and $K = 0.5$. The current angular momentum of the Earth–Moon system is positive. Therefore, an angular momentum equal to $K_{sEM}$ for final condensation mass $m_f \approx 0.15(M_E + M_M)$ may be acquired at $k_H = 1$ and $K = 0.9$ with any contribution of collisions of the considered parental condensation with small objects (i.e., with any contribution of the collision of two large condensations) to the angular momentum of the parental condensation that contracted and formed embryos of the Moon and the Earth. The above estimates of the condensation mass needed to form the Earth–Moon system may be reduced if one takes the increase in $K_s$ during the growth of embryos of the Moon and the Earth into account.

It is theoretically possible that the angular momentum of a condensation needed to form the Earth–Moon system was acquired through the accumulation of small objects by a condensation with final mass $m_f > 0.15 M_E$, but we believe that the collision of large condensations produced the dominant contribution to the angular momentum of the parental condensation. If this were not the case, the parental condensations of Venus and Mars could also acquire angular momenta sufficient to form large satellites. It is likely that the Earth differed from other terrestrial planets in that the condensations contracting to form embryos of these planets did not collide with massive condensations and thus did not acquire angular momenta required to form massive satellites. The collision of condensations producing a condensation with an angular momentum sufficient to form an embryo of a planet with a massive satellite could occur only in the evolutionary history of the Earth.

The accumulation of small objects fails to account for the current tilts (~23°–25°) of the rotation axes of the Earth and Mars, since the tilt of the rotation axis of a condensation (and a solid planetary embryo) is near-zero in the case of accumulation of small objects. The larger the contribution of small objects to the formation of the parental condensation for the Earth–Moon system, the lower the possible masses of condensations in the primary collision. It would be instructive to determine, using the model of condensation formation, the maximum masses of two condensations that are both located at a distance of ~1 AU from the Sun and differ in mass by no more than an order of magnitude.

## 2. GROWTH OF SOLID EMBRYOS OF THE EARTH AND THE MOON

### 2.1. Relative Variations of Masses of Embryos of the Earth and the Moon When Accumulating Planetesimals

In the present subsection, the relative growth of embryos of the Earth and the Moon while accumulating planetesimals (or any other objects falling within the Hill sphere of the Earth) is compared. The increase in mass of a celestial body is proportional to the square of the effective radius $r_{eff}$ (area of a circle with a radius equal to effective radius $r_{eff}$). Effective radius $r_{eff}$ is the impact parameter at which a planet (celestial body) is reached. It is written as

$$r_{ef} = r \cdot (1 + (v_p/v_r)^2)^{1/2}, \quad (5)$$

where $v_p$ is the parabolic velocity on the planetary surface and $v_r$ is the relative velocity at infinity (Okhotsimskii, 1968, pp. 36–37). If $v_r > v_p$ (e.g., for comets infalling onto the Earth from highly eccentric orbits), $r_{eff}$ is close to $r$.

If the relative velocities are low and $(v_p/v_r)^2$ is much larger than 1, $r_{eff}$ is close to $r(v_p/v_r)$, where $v_p = (2Gm/r)^{1/2}$ and $m$ is the mass of a planet with radius $r$. Then, $r_{eff}$ is close to $r(v_p/v_r) = r(2Gm/r)^{1/2}/v_r = (8G\pi\rho/3)^{1/2} r^2/v_r$, where $m = (4/3)\pi\rho r^3$ and $\rho$ is the density of a planet. Therefore, in the case of low relative velocities (with $r_{eff}^2$ proportional to $\rho r^4$), ratio $dm/dt$ is proportional to $\rho r^4$ (i.e., to $\rho^{-1/3} m^{4/3}$, since $r$ is proportional to $(m/\rho)^{1/3}$ and $r^4$ is proportional to $(m/\rho)^{4/3}$). Thus, $dm/(\rho^{-1/3} m^{4/3}) = cdt$, where $c$ is mass-independent. Having integrated relation $\rho_E^{1/3} \cdot m_E^{-4/3} dm_E = \rho_M^{1/3} m_M^{-4/3} dm_M$, we obtain

$$m_{Mo}^{-1/3} = m_M^{-1/3} + k_2 m_{Eo}^{-1/3} - k_2 m_E^{-1/3}, \quad (6)$$

where $k_2 = k_d^{1/3} = (\rho_E/\rho_M)^{1/3}$, $k_d = \rho_E/\rho_M$ is the ratio of the density of the growing Earth with mass $m_E$ to the density of the growing Moon with mass $m_M$ ($k_d \approx 1.65$ for the current densities of the Earth and the Moon), and $m_{Eo}$ and $m_{Mo}$ are the initial masses of embryos of the Earth and the Moon, respectively. If $m_M = 0.0123 m_E$, $m_{Eo} = 0.1 m_E$, and $m_E = M_E$, relation (6) holds true at $k_d = 1$ and $m_{Mo} = 0.00605 M_E$ and at $k_d = 1.65$ and $m_{Mo} = 0.0054 M_E$.

The above estimates of the relative increase in masses of embryos were obtained in the model where the distance between these embryos is larger than the Hill sphere of the Earth embryo. Owing to the gravitational influence of the Earth embryo, the probability of a collision between a planetesimal and the Moon embryo increases as the distance between the embryos gets shorter. This factor enhances the relative increase in mass of the Moon embryo in the case of a short distance between the embryos and makes the conclusions regarding the iron abundance in growing embryos (see below) even more well-founded.

If $r_{eff}$ is close to $r$, $dm/dt$ is proportional to $r^2$ (i.e., to $\rho^{-2/3}m^{2/3}$, since $r$ is proportional to $(m/\rho)^{1/3}$ and $r^2$ is proportional to $(m/\rho)^{2/3}$). Thus, $dm/(\rho^{-2/3}m^{2/3}) = c_2 dt$, where $c_2$ is mass-independent. We consider relation $\rho_E^{2/3}m_E^{-2/3}dm_E = \rho_M^{2/3}m_M^{-2/3}dm_M$. The following is $m_{Mo}^{1/3} = m_M^{1/3} + k_1 m_{Eo}^{1/3} - k_1 m_E^{1/3}$, where $k_1 = k_d^{2/3}$.

If $r_{eff}$ is proportional to $r^2$, the embryo of the Earth grows faster than the Moon embryo. For example, the mass of the Moon embryo increases by a factor of 2 at $k_d = 1$ and 2.3 at $k_d = 1.65$, while the mass of the Earth embryo increases by a factor of 10. It is conceivable that the effective radii of the proto-Earth and the proto-Moon were proportional to $r$ at sufficiently high eccentricities of planetesimal orbits. The growth of $m_M/m_{Mo}$ then outpaces the $m_E/m_{Eo}$ growth. The above models demonstrate that the relative growth of embryos of the Earth and the Moon depended to a considerable extent on the eccentricities of orbits of infalling planetesimals.

## 2.2. Increase in the Angular Momentum of Embryos of the Earth–Moon System

The total angular momentum $K_\Sigma$ of embryos of the Earth and the Moon increased as they grew. This momentum included the angular momenta of embryos relative to their centers of mass and angular momentum $K_{ME}$ of embryos of the Earth and the Moon relative to their common center of mass. The growth of $K_\Sigma$ was influenced by many factors. The current angular momentum of the Earth and the Moon relative to their centers of mass amounts to 17% of the total angular momentum of the system (Barr, 2016).

According to Eq. (2), the angular momentum of a condensation with mass $m_c$ formed in a collision of two identical condensations which moved before their collision in circular heliocentric orbits is proportional to $m_c^{5/3}$. It follows from Eq. (4) that the growth of the angular momentum of a planet with mass $m_{pl}$ under the infall of bodies with velocities proportional to the parabolic velocity on the planetary surface is proportional to $m_{pl}^{5/3}$. With such proportionality relations ($m_c^{5/3}$ and $m_{pl}^{5/3}$), the angular momentum of a planet grown from an embryo with mass $m_{co}$ should not depend on $m_{co}$.

Angular momentum $K_{ME}$ of embryos of the Earth and the Moon relative to their common center of mass may be written as $K_{ME} = (r_{EM}G)^{1/2}m_M m_E(m_E + m_M)^{-1/2}$, where $r_{EM}$ is the distance between the embryos. At $r_{EM}=\text{const}$ and $m_E \gg m_M$, $K_{ME}$ is proportional to $m_M \cdot m_E^{1/2}$. It was obtained above at $k_d = 1.65$ and $r_{eff}$ proportional to $r^2$ that the mass of the Moon embryo increases by a factor of 2.3, while the mass of the Earth embryo increases by a factor of 10. Then, if $m_E$ increases 10 times, $m_M \cdot m_E^{1/2}$ increases by a factor of $7.6 \approx 10^{0.88}$. Let us assume that this growth of $m_M \cdot m_E^{1/2}$ by a factor of $m_E^{0.88}$ is also taken place at other $m_E$ values.

It was already noted that $K_\Sigma$ may be equal to current angular momentum $K_{\Sigma EM}$ of the Earth–Moon system at $m_{EM} = m_E + m_M \approx 0.1(M_E + M_M)$. Let us consider a model with $K_{ME}$ being the major part of angular momentum $K_\Sigma$ of the system. According to Eq. (2), $K_\Sigma = (\alpha/0.1)^{5/3} K_{\Sigma EM}$ for the initial embryos with total mass $\alpha(M_E + M_M)$. Assuming that $K_\Sigma$ increased by a factor of $\alpha^{-0.88}$ and became equal to $K_{\Sigma EM}$ in during planetesimal accumulation and increase in the mass of embryos from $\alpha(M_E + M_M)$ to $M_E + M_M$, we obtain the following: $(\alpha/0.1)^{5/3}\alpha^{-0.88} = 1$. From this relationship we get $\alpha \approx 0.0078$. Therefore, the initial mass of a condensation producing embryos of the Earth and the Moon could theoretically be lower than $0.01(M_E + M_M)$. This estimate was obtained while modeling the increases in the masses of embryos without regard for the sources of infalling bodies and remains valid in the case of infall of matter ejected from the Earth embryo onto the Moon embryo. Since a fraction of the matter escapes in the process of condensation contraction and the distance between the initial embryos was originally shorter than the Earth–Moon distance, the larger estimate of the mass of the initial condensation is more probable. The estimates of mass of the original parental condensation increase when one takes into account the fact that the infall of planetesimals onto the embryos primarily occurred at an interembryo distance shorter than the current Earth–Moon distance. On the other hand, it is often assumed that the axial rotation period of the Earth was shorter in the past; therefore, its angular momentum was larger. The consideration of this factor reduces the contribution of the $K_{ME}$ growth induced by the accumulation of planetesimals by the embryos.

## 2.3. Variation of the Iron Abundance in the Growing Embryos of the Earth and the Moon under the Infall of Planetesimals

According to (Galimov et al., 2005; Galimov, 2011; Galimov and Krivtsov, 2012), the original embryos of the Earth and the Moon produced as a result of condensation contraction contained a relatively small amount of iron, and the Earth, which grew faster due to the accumulation of dust, acquired more iron than the Moon. In the present subsection, we consider the growth of iron-depleted embryos of the Earth and the Moon induced exclusively by the infall of planetesimals. A simple supplementary model may be used to estimate the maximum increase in mass $m_M$ of the Moon embryo. According to this model, the initial embryos contained no iron, while the infalling matter contained 33% Fe. The abundance of iron on the Moon is then $0.33(1 - m_r)$, where $m_r$ is the ratio of the initial mass of the Moon embryo to the current mass of the Moon. Assuming that the current iron abundance on the Moon is 8% (Barr, 2016), we obtain $m_r = 0.76$ and a 1.3-fold growth of the Moon embryo from the $0.33(1 - m_r) = 0.08$ relation. This estimate agrees with those reported in (Galimov and Krivtsov, 2012), where the Moon embryo grew by a factor of 1.31 as the mass of the Earth embryo increased by a factor of $k_E = 26.2$. With these estimates of Galimov and Krivtsov, increment $dm$ of embryo mass $m$ is proportional to $m^2$ (i.e., $r_{eff}$ is proportional to $r^3$). In the considered auxiliary model, the iron abundance on the Earth estimated at $k_E = 26.2$ is $0.33(1 - 1/26.2) = 0.317$, which is close to the actual value of 32%. In view of Galimov and Krivtsov (2012), the concentration of iron in dust particles after evaporation of 40% of their matter is close to the iron abundance on the Moon. If this is the case, the current iron abundance is reproduced only if the Moon embryo did not accumulate any planetesimals at all, and it follows from the $0.33(1 - m_{Eo}/M_E) + 0.08 m_{Eo}/M_E = 0.32$ relation that mass $m_{Eo}$ of the initial Earth embryo with 8% Fe was $0.04 M_E$. With these estimates, the mass of the Earth embryo grows by a factor of 25 while the mass of the Moon remains unchanged. In order to obtain the current iron abundance on the Earth and the Moon with a nonzero iron concentration in the initial embryos, increment $dm$ of the embryo mass should be proportional to $m^p$, where $p > 2$.

The calculations of increases in the masses of embryos of the Earth and the Moon in (Galimov and Krivtsov, 2012; Vasil'ev et al., 2011) were performed in the model where the particle velocities were zero at the boundaries of a cylinder in the Hill sphere of the larger embryo. The embryos formed in the process of con-traction of a condensation with a radius of 5.5 Earth radii and a mass of $1.023 M_E$ grew by accumulating iron-enriched material in the outer region of the condensation that remained within the Hill sphere after the formation of the initial embryos. Thus, the total mass of material within the Hill sphere in the model of Galimov et al. should be considerably larger than $M_E$ (with the inner and outer regions of the condensation both having a mass on the order of $M_E$). In this model, particles forming a condensation with a radius of 5.5 Earth radii were depleted in iron, while the other particles in the Hill sphere were iron-enriched. According to (Galimov and Krivtsov, 2012), the total mass of initial embryos needed to reproduce the current iron abundance on the Earth and the Moon is $0.047 M_E = M_E/26.2 + M_E/(1.31 \times 81.3)$. The calculations of con-traction of the condensation with a radius of 5.5 Earth radii were performed in this study for a condensation mass of $1.023 M_E$. Galimov et al. have not indicated where 95% of iron-depleted material from the inner part of the condensation had gone before the embryos started growing by accumulating matter from the outer part of the condensation.

Let us discuss several modifications of the model of Galimov et al. that may rectify some of the above drawbacks. In order to form small embryos, one may calculate the contraction of a less massive condensation (with a mass considerably smaller than $M_E$) so as to obtain embryos that incorporate almost all iron-depleted material. In our view, the calculations of migration of particles toward embryos from the cylinder surface within the Hill sphere of the Earth embryo in (Galimov and Krivtsov, 2012; Vasil'ev et al., 2011) may also be interpreted as the inflow of matter from outside the Hill sphere. However, in the case of matter inflow from outside the Hill sphere, the considered model with zero relative velocities is hardly relevant. Owing to the condensation rotation, the particle velocities were nonzero even inside the condensation. If the velocities were zero, the particles would very quickly (according to (Wahlberg Jansson and Johansen, 2014), the free-fall time is 25 years) reach the center of the condensation and would not "wait" for the embryos to form in the hot inner part of the condensation, where particle evaporation alone took tens of thousands of years (Galimov and Krivtsov, 2012). The higher the relative velocities of particles on their entry into the Hill sphere, the smaller the difference in the relative growth of mass for embryos of the Earth and the Moon (the smaller the relative growth of the Earth embryo). The higher (compared to planetesimals) relative probability of infall of particles onto the Earth embryo (compared to the Moon embryo) obtained in (Galimov and Krivtsov, 2012; Vasil'ev et al., 2011) is established by zero relative particle velocities. The relative probability should be even higher to ensure the abovementioned $dm$ growth proportional to $m^p$ at $p > 2$. For this growth to be sustained, one may assume that the ejection of matter from the surface of the Moon embryo in the process of infall of planetesimals almost halted its growth.

It is also not implausible that the contraction of the central part of the condensation, which resulted in the formation of embryos, was accompanied by the contraction of the entire condensation with the size of the Hill sphere. Galimov and Krivtsov (2012) have noted that the rapid contraction of the central ($r < 5.5 r_E$) part of the condensation was hindered by the high temperature in this region. The outer ($r > 5.5 r_E$) part was assumed to be cooler. Thus, it seems that the temperature in the outer part of the condensation should be less of an obstacle to contraction than in the central part, and a question then arises as to how the relatively cool outer part of the condensation could survive in the model of Galimov et al. for 50 million years after the formation of the Solar System.

No studies into the formation of condensations with masses no lower than the Earth mass in the region of terrestrial planets have been published yet. Therefore, the question of the possibility of the formation of massive condensations considered in (Galimov and Krivtsov, 2012; Vasil'ev et al., 2011) remains open. A fraction of the condensation material, which was not incorporated into the embryos, could leave the Hill sphere during condensation contraction, thus increasing the mass of the parental condensation. A consider-able amount of matter possibly falling onto embryos of the Earth and the Moon could be present outside the Hill sphere of the parental condensation.

The objects (e.g., planetesimals) falling onto embryos of the Earth and the Moon in the model considered in Subsection 2.1 originated from outside the Hill sphere. In the calculations of Ipatov (1993, 2000), the mean eccentricities of planetesimal orbits in the feeding zone of terrestrial planets exceeded 0.2 (and then 0.3) at certain stages of evolution. At such eccentricities, parameter $p$ was below 4/3. The dependence of $r_{eff}$ on $r$ may vary from $r$ to $r^2$ (depending on eccentricities) during accumulation of planetesimals entering the Hill spheres of embryos from the outside. In these extreme cases, d$m$ is proportional to $m^{2/3}$ or $m^{4/3}$, respectively, and the ratio of masses of embryos of the Moon and the Earth ($m_M/m_E$) increases faster with $m_E$ than in (Galimov and Krivtsov, 2012; Vasil'ev et al., 2011). If $r_{ef}$ is proportional to $r^2$ (the case of low-eccentricity planetesimal orbits), a 1.3-fold increase in the mass of the Moon embryo (to the current mass of the Moon) corresponds, according to Eq. (6), to a 2.4-fold and 2.7-fold (to $M_E$) increase in the mass of the Earth embryo at $k_d = 1.65$ and $k_d = 1$, respectively. In the model with no iron in the initial embryos, the abundance of iron on the Earth does not exceed 0.33(1–1/2.7) ≈ 0.21 (i.e., is lower than the current level). With this $m_E$ growth, the concentration of iron in planetesimals should be no lower than 0.32/(1 − 1/2.7) = 0.32 × 2.7/1.7 ≈ 0.5 (which is infeasible) in order to reproduce the current iron abundance (32%) on the Earth. If the concentration of iron in the initial embryos is nonzero, the current iron abundance on the Moon is attained as the mass of the Moon embryo increases by a factor smaller than 1.3. Even at an iron concentration of 8% in the Earth embryo, the iron abundance does not exceed 0.24 after the mass of the embryo increases 2.4–2.7-fold.

The above estimates suggest that it is hard to repro-duce the current iron abundances on the Earth and the Moon at any initial Fe concentrations in the initial embryos growing exclusively through the accumulation of solid planetesimals (without matter ejection on impacts). In order to obtain the current iron abundances at a nonzero (although low) iron concentration in the initial embryos, the increase d$m$ in the embryo mass should be proportional to $m^p$, where $p > 2$. Is it possible (excluding the case of matter ejection from embryos)? Parameter $p$ for the motion of solid bodies in a gas-free medium does not exceed 4/3. If the gas drag is taken into account and/or dust particles are considered, the value of $p$ may vary.

Levison et al. (2015) believed that planetesimals grew immediately after their formation by accumulating pebble-sized bodies in the presence of gas. The growth of embryos in the presence of gas may be examined using the formulas presented in (Ormel and Klahr, 2010; Levison et al., 2015; Chambers, 2017). According to Levison et al. (2015), the cross section for the capture of bodies by an embryo is given by

$$S = 4\pi G \cdot m_e t_s v_{rel}^{-1} \exp^{-\xi}, \qquad (7)$$

where $\xi = 2[t_s v_{rel}^3 /(4Gm_e)]^{0.65}$, $m_e$ is the mass of the embryo, $v_{rel}$ is the velocity of the body relative to the embryo, and $t_s$ is the stopping time due to aerodynamic drag. It follows from Eq. (7) that the relative increase in the mass of embryos of the Earth and the Moon in unit time is proportional to $m_{rEM} \exp^{-\zeta}$ (but depends also on the values of $t_s$ and $v_{rel}$, which differ from one embryo to the other), where $\zeta = m_{rEM}^{-0.65}$, and $m_{rEM}$ is the ratio of masses of embryos. At $m_{rEM} = 10$ and 30, $m_{rEM} \exp^{-\zeta}$ assumes a value of 8 and 27, respectively. Note that at $m_{rEM} > 5$, $0.7 < \exp^{-\zeta} < 1$ and $m_{rEM} \exp^{-\zeta}$ is close to $m_{rEM}$. The relative increase in the mass of the Earth embryo is then no larger than in the model of infall of solid bodies in the case of low relative velocities with $r_{eff}^2$ proportional to $m^{4/3}$, which was considered above (in the beginning of Subsection 2.1).

The results reported in (Hughes and Boley, 2017) demonstrate that the influence of gas on the effective cross section of an embryo depends on the size of infalling objects, the size and the density of the embryo, and on the distance from the embryo to the Sun. In the calculations performed in this study for a distance of 1 AU from the Sun, objects ~0.3 cm in size were captured most efficiently in a gas. The ratio of the effective radius to the radius of the object for smaller particles in a gas may be considerably lower than the ratio in formula (5) for a gas-free medium, since such particles in a gas flow around the embryo. Table 2 from (Hughes and Boley, 2017) shows that the increase in the embryo mass attributable to the accumulation of larger objects differs by a factor of no more than two from the mass gain at such object sizes as ensure the maximum growth. Additional studies, which would include, among other things, the mass distribution of particles and other small objects, are needed in order to draw certain conclusions regarding the relative growth of embryos of the Earth and the Moon through the accumulation of small objects moving in a gas. If we consider the infall of a large number of relatively small bodies onto these embryos, they could acquire matter with roughly the same isotopic composition.

### 2.4. Growth of the Embryo of the Moon Induced by the Infall of Matter Ejected from the Embryo of the Earth

When bombarded by planetesimals, the embryo of the Moon, which was originally located closer to the embryo of the Earth, could grow primarily by accumulating iron-depleted matter of the crust and the mantle of the Earth ejected from the surface of the Earth embryo in its collisions with planetesimals and smaller bodies. This source of growth of the Moon embryo does not impose any significant constraints on the initial masses of embryos of the Moon and the Earth and their parental condensation. This model does not require the initial embryos of the Moon and the Earth to be depleted in iron. The ratio of contributions of infalling planetesimals and the matter ejected from the Earth embryo to the growth of the Moon embryo depends on the results of collisions of planetesimals with the embryos and on the distance between the embryos. Since the velocities of collisions with the Moon embryo are lower for the bodies ejected from the Earth embryo, the probability of their capture is higher than the probability of capture of directly infalling bodies.

If the iron abundance in the initial Moon embryo and in planetesimals was 0.33 and the iron abundance in the crust of the Earth and on the Moon was 0.05 and 0.08, respectively, fraction $k_E$ of matter of the Earth crust in the Moon should be ~0.9 (this follows from relation $0.05 k_E + 0.33(1 - k_E) = 0.08$). Therefore, in order to reproduce the current iron abundance on the Moon, the amount of matter ejected from the Earth embryo and accumulated by the Moon embryo should be an order of magnitude larger than the sum of the total mass of planetesimals falling onto the Moon embryo and the initial mass of the Moon embryo formed from the parental condensation (if the iron abundance in the initial embryo was the same as in planetesimals). The estimated fraction of matter of the Earth's crust in the Moon decreases as the mass of the Moon embryo formed in the process of condensation contraction increases and as the concentration of iron in it decreases.

The considered approach with the initial embryos of the Moon and the Earth forming from a common parental condensation differs considerably from the one used in (Ringwood, 1989; Vityazev and Pechernikova, 1996; Gorkavyi, 2004, 2007; Svetsov et al., 2012; Citron et al., 2014; Rufu and Aharonson, 2015, 2017), where the formation and growth of the Moon embryo primarily through the accumulation of matter of the Earth's crust ejected from the Earth embryo in multiple collisions with bodies from the protoplanetary disk was considered. In the model used in the present study, both embryos formed from the same condensation. The subsequent growth of embryos of the Moon and the Earth formed in the process of contraction of the parental condensation was the same as in the multi-impact model. Matter incorporated into the Moon embryo could be ejected from the Earth in multiple collisions between planetesimals (and smaller bodies) and the Earth, while Rufu and Aharonson (2017) considered only ~20 massive collisions.

Objects ejected from the Earth embryo in collisions with other objects were more likely to be incorporated into the large Moon embryo than to merge with similar low-mass objects. Therefore, the presence of the large Moon embryo formed during the contraction of a condensation made the formation of a larger (compared to the case of formation exclusively from matter ejected from the Earth) satellite of the Earth possible. This is the likely reason why Venus lacks a satellite. The parental condensations of embryos of Venus, Mars, and Mercury did not acquire an angular momentum sufficient to form a large satellite. Planetesimals falling onto Venus and the Earth had approximately the same mass and velocity distributions. Matter was also ejected from the surface of Venus after its collisions with these planetesimals, but no satellite was formed from this matter.

The masses of impactors in the above-cited studies focused on the multi-impact model did not exceed $0.1 M_E$. Collisions of the proto-Earth with impactors with masses below $0.1 M_E$ were considered in the megaimpact model even before the calculations within the multi-impact model. The needed angular momentum at such masses may be acquired in the megaimpact model only in a grazing collision with the protolunar disk formed primarily from the impactor material and containing a considerable amount of iron (Canup, 2012). Trying to reproduce the composition of the Moon within the megaimpact model, (Cuk and Stewart, 2012) have considered an impactor with a mass below $0.1 M_E$, an almost head-on collision, and a pre-collision axial rotation period of the proto-Earth of 2.3 h, while (Canup, 2012) has modeled an impactor with a mass of $0.4 M_E - 0.5 M_E$. Since the formation of the Moon in a single collision is not required in the multi-impact model, a rapidly rotating proto-Earth or a very massive impactor are also not needed to reproduce the composition of the Moon.

Two condensations, which had collided and formed the condensation that contracted to produce embryos of the Moon and the Earth, could move in different planes around the Sun prior to the collision. Therefore, the orbital plane of the Moon embryo was not necessarily aligned with the ecliptic plane. Note that the angle between the Moon's orbit and the ecliptic plane is 5.1°.

A single collision (or a series of collisions) between the solid Earth embryo and a massive object or an additional collision of condensations are needed for the rotation axis of the Earth to acquire the current tilt. A collision of condensations could contribute to this tilt if at the time when the parental condensation had split into two components, the radius of the condensation producing the Earth embryo was, at the moment of this collision, shorter than the semi-major axis of the orbit of the condensation that produced the Moon embryo and moved around the Earth embryo.

Ipatov (1981b) has demonstrated that if the vector of the angular momentum of a planet with respect to its center of mass is perpendicular to its orbital plane prior to the collision with an impactor and the vector of gain in the angular momentum in the collision is perpendicular to the vector of this momentum, the ratio of masses of the impactor and the planet is $m_I/m_{pl} \approx 2.5 r_{pl} \chi \tan I / (\alpha v_{par} T_{pl}(1 + \tan^2 I)^{1/2})$, where $r_{pl}$ is the radius of the planet, $T_{pl}$ is the period of its axial rotation, $\alpha v_{par}$ is the tangential component of the collision velocity, $v_{par}$ is the parabolic velocity on the planetary surface, and $\tan I$ is the tangent of angle $I$ between the axis of rotation of the planet after the collision and the normal to the orbital plane of the planet. At $\chi = \alpha = 1$, $T_{pl} = 24^h$, and $I = 23.44°$, we obtain $m_I/m_{pl} \approx 0.0065$. Thus, the current tilt of the rotation axis of the Earth could be acquired in a collision with an impactor with a mass of ~$0.01 M_E$.

The considered model may also be applicable to the formation of an exoplanet with a large satellite.

## CONCLUSIONS

The angular momentum of a parental condensation needed to form embryos of the Earth and the Moon could mostly be acquired in a collision of two rarefied condensations producing the parental condensation. The angular momentum of the Earth–Moon system could be acquired in a collision of two condensations in circular heliocentric orbits with their total mass being no lower than the mass of Mars. With the subsequent growth of embryos of the Moon and the Earth taken into account, the total mass of embryos, which had formed in the process of contraction of the parental condensation, needed to reach the current angular momentum of the Earth–Moon system could be below 0.01 of the Earth's mass. For the low lunar iron abundance to be reproduced with the growth of originally iron-depleted embryos of the Moon and the Earth just by the accretion of planetesimals, the mass of the lunar embryo should have increased by a factor of 1.3 at the most. The maximum increase in the mass of the Earth embryo due to the accumulation of planetesimals in a gas-free medium is then threefold, and the current terrestrial iron abundance is not attained. If the embryos are assumed to have grown just by accumulating solid planetesimals (without the ejection of matter from the embryos), it is hard to reproduce the current lunar and terrestrial iron abundances at any initial abundance in the embryos. In order to obtain the current iron abundance on the Earth and the Moon with a certain concentration of iron in the initial embryos, increment d$m$ of the embryo mass should be proportional to $m^p$, where $p \geq 2$. Parameter $p$ for the motion of solid bodies in a gas-free medium does not exceed 4/3. In order to reproduce the current iron abundance on the Moon, the amount of matter ejected from the Earth embryo and accumulated by the Moon embryo should be an order of magnitude larger than the sum of the total mass of planetesimals falling onto the Moon embryo and the initial mass of the Moon embryo formed from the parental condensation (if the iron abundance in the initial embryo was the same as in planetesimals). The greater part of matter incorporated into the Moon embryo could be ejected from the Earth in its multiple collisions with planetesimals (and smaller bodies).

## ACKNOWLEDGMENTS

This study was supported financially by the Russian Science Foundation, project no. 17-17-01279 (formation and growth of embryos of the Earth and the Moon), and program no. 28 of the Presidium of the Russian Academy of Sciences under state assignment no. 00137-2018-0033 to the Vernadsky Institute of Geochemistry and Analytical Chemistry (research into the angular momentum of colliding celestial bodies).

The author wishes to thank Academician M. Ya. Marov for his interest in the research and helpful suggestions, which have improved the manuscript, and Academician E.M. Galimov for a detailed account of his model.